\documentclass[11pt,journal,compsoc,onecolumn]{IEEEtran}
\usepackage{balance}  

\usepackage{setspace}

\usepackage{amsmath}
\usepackage{amsfonts}
\usepackage{amssymb}
\usepackage{amsthm}
\usepackage{balance}
\usepackage{url}

\usepackage[ruled,vlined]{algorithm2e}
\renewcommand{\CommentSty}[1]{\textnormal{#1}}
\DontPrintSemicolon
\SetKwComment{tcp}{$\triangleright$ }{}
\SetVlineSkip{0cm}
\SetKwInOut{Output}{Output}
\SetKwInOut{Input}{Input}

\usepackage{graphicx}
\usepackage{subfigure}

\theoremstyle{definition}

\usepackage{color}

\usepackage{multicol}
\usepackage{multirow}

\usepackage{footnote}
\makesavenoteenv{tabular}
\makesavenoteenv{table}

\usepackage{xargs}
\usepackage{xspace}
\usepackage[pdftex,dvipsnames]{xcolor}  

\definecolor{midnight}{rgb}{0,0.094,0.533}

\usepackage[colorlinks=true,%
   linkcolor=red,%
   citecolor=midnight,%
   urlcolor=blue,%
   bookmarks=false]{hyperref}

\usepackage[sort,compress]{cite}

\usepackage{pifont}

\usepackage[ruled,vlined]{algorithm2e}
\renewcommand{\CommentSty}[1]{\textnormal{#1}}
\DontPrintSemicolon
\SetKwComment{tcp}{$\triangleright$ }{}
\SetVlineSkip{0cm}
\SetKwInOut{Output}{Output}
\SetKwInOut{Input}{Input}
\makeatletter
\newcommand{\removelatexerror}{\let\@latex@error\@gobble}
\makeatother

\usepackage{booktabs}   
\usepackage{subfigure} 

\usepackage{listings}
\usepackage{colortbl}
\usepackage{color}

\definecolor{dkgreen}{rgb}{0,0.6,0}
\definecolor{dred}{rgb}{0.545,0,0}
\definecolor{dblue}{rgb}{0,0,0.545}
\definecolor{lgrey}{rgb}{0.9,0.9,0.9}
\definecolor{gray}{rgb}{0.4,0.4,0.4}
\definecolor{darkblue}{rgb}{0.0,0.0,0.6}
\definecolor{orange}{HTML}{ab388e}
\definecolor{ddblue}{HTML}{4472c2}

\definecolor{red}{rgb}{0.8,0,0}
\definecolor{green}{rgb}{0,0.6,0}

\usepackage{epigraph}
\usepackage{etoolbox}
\makeatletter
\patchcmd{\epigraph}{\@epitext{#1}}{\itshape\@epitext{#1}}{}{}
\makeatother

\lstdefinelanguage{pgabb}{
      basicstyle=\footnotesize \ttfamily \color{black},
      breakatwhitespace=false,
      breaklines=true,
      captionpos=b,
      commentstyle=\color{dkgreen},
      deletekeywords={...},
      escapeinside={\%*}{*)},
      language=C++,
      keywordstyle=\color{orange},
      morekeywords={auto, pgabb,mod, max, min, functor, in, size_t,
                    B, T, L, V, BlockList,
                    for_dev, for_host, reduce_dev, reduce_host},
      emph={start,stop, true, false,
            Vertices, Edges, GetInterval, Iteration, Add, Attribute, Sync,
            AssignAttribute,
            CAS, CASS},
      emphstyle=\color{ddblue},
      identifierstyle=\color{black},
      stringstyle=\color{blue},
      numbers=left,
      numbersep=5pt,
      numberstyle=\tiny\color{black},
      rulecolor=\color{black},
      showspaces=false,
      showstringspaces=false,
      showtabs=false,
      stepnumber=1,
      tabsize=5,
      title=\lstname,
}
\usepackage{newverbs}
\newverbcommand{\emphverb}{\color{ddblue}}{}
\newverbcommand{\kwverb}{\color{orange}}{}

\newcommand{\pgabb}{{PGAbB}\xspace}
\newcommand{\hostkernel}{$\mathcal{K}_H$\xspace}
\newcommand{\devkernel}{$\mathcal{K}_D$\xspace}
\newcommand{\patterngen}{$\mathcal{P}_G$\xspace}
\newcommand{\patterncus}{$\mathcal{P}_C$\xspace}
\newcommand{\pre}{$\mathcal{I}_B$\xspace}
\newcommand{\post}{$\mathcal{I}_A$\xspace}
\newcommand{\est}{$\mathcal{E}$\xspace}

\usepackage{multirow}
\usepackage{array}
\newcolumntype{L}[1]{>{\raggedright\let\newline\\\arraybackslash\hspace{0pt}}m{#1}}
\newcolumntype{C}[1]{>{\centering\let\newline\\\arraybackslash\hspace{0pt}}m{#1}}
\newcolumntype{R}[1]{>{\raggedleft\let\newline\\\arraybackslash\hspace{0pt}}m{#1}}

\usepackage{xargs}
\usepackage{xspace}
\usepackage[colorinlistoftodos,prependcaption,textsize=tiny]{todonotes}

\newcommandx{\uvc}[2][1=]{\todo[color=magenta!50,#1]{\sf \textbf{\"Umit:} #2}\xspace}

\newcommandx{\ay}[2][1=]{\todo[color=green!50,#1]{\sf \textbf{Apo:} #2}\xspace}
\usepackage{outlines}

\begin{document}

\title{PGAbB: A Block-Based Graph Processing Framework for Heterogeneous Platforms}

\author{
    Abdurrahman Ya\c{s}ar~\textsuperscript{*},
    Sivasankaran Rajamanickam~\IEEEmembership{(Member,~IEEE)},
    Jonathan W. Berry, and\\
    \"{U}mit V. \c{C}ataly\"{u}rek~\IEEEmembership{(Fellow,~IEEE)}~\textsuperscript{*},
    \IEEEcompsocitemizethanks{\IEEEcompsocthanksitem Ya\c{s}ar, is with Nvidia, Santa Clara, CA.
    E-mail: ayasar@nvidia.com
    \IEEEcompsocthanksitem \c{C}ataly\"{u}rek
    is with the School of Computational Science and Engineering at Georgia Institute
    of Technology. E-mail: umit@gatech.edu
    \IEEEcompsocthanksitem Rajamanickam, and Berry are with
    the Center for Computing Research at Sandia National Laboratories, Albuquerque,
    NM.\protect\\
    E-mail: \{srajama,jberry\}@sandia.gov
    }
    
    \thanks{Manuscript received xx xx, xx; revised xx xx, xx.}}
    
    \IEEEtitleabstractindextext{%
\begin{abstract}
Designing flexible graph kernels that can run well on various
platforms is a crucial research problem due to the frequent usage of graphs
for modeling data and recent architectural advances and variety. In this
work, we propose a novel graph processing framework,
\pgabb (Parallel Graph Algorithms by Blocks), for
modern shared-memory heterogeneous platforms. Our
framework implements a block-based programming model. This allows a user to
express a graph algorithm using kernels that operate on subgraphs.
\pgabb support graph computations that fit in host DRAM but not in
GPU device memory, and provides simple but effective scheduling techniques to
schedule computations to all available
resources in a heterogeneous architecture.
We have demonstrated that one can easily implement a diverse set of graph algorithms in
our framework by developing five algorithms.
Our experimental results show that \pgabb implementations
achieve better or competitive performance compared to hand-optimized
implementations. Based on our experiments on five graph algorithms and
forty-four graphs, in the median, \pgabb achieves 1.6, 1.6,
5.7, 3.4, 4.5, and 2.4 times better performance than
GAPBS, Galois, Ligra, LAGraph Galois-GPU, and Gunrock graph
processing systems, respectively.
\end{abstract}

}
  
\maketitle

\begingroup\renewcommand\thefootnote{*}
\footnotetext{This publication describes work performed at the Georgia Institute of Technology and is not associated with Nvidia and Amazon.}
\endgroup

\IEEEdisplaynontitleabstractindextext
\IEEEpeerreviewmaketitle

\section{Introduction}
\label{sec:intro}
\epigraph{\hfill ``Mobilis in Mobili'', --- \textup{Jules Verne}}{}
\vspace{-1.2em}

Graphs are one of the dominant data structures to model irregular and complex
data. Hence, graph
analysis is crucial for many data analytics applications.
High-performance parallel graph
processing has been an active research area for decades, and has become
even more active in recent years as computer architectures have evolved.
Reducing the running times of graph algorithms lowers overall costs and eases
pressure on the
environment, while easy implementation of efficient algorithms increases
productivity. However, developing high-performance parallel graph algorithms
is a challenging task. Several well-known hardware and
software-related challenges arise~\cite{Lumsdaine07-PPL}.
Graph algorithms have random data access patterns,
limiting their locality. Furthermore, these computations are memory
and network bound rather than compute bound since the ratio of data access to
computation is high, and data re-use is low. Finally, graph algorithms
are data-driven and graph datasets tend to be highly skewed by vertex degree.
This causes a significant workload
imbalance between computational loads. An efficient parallel graph
algorithm must address these challenges.
Different architectures are ideal for addressing different kinds
of problems. Therefore,
heterogeneous systems can be very beneficial in addressing wide
range of challenges instead of solely depending on an architecture.

{\bf Limitation of state-of-art approaches:}
For a long time, distributed systems were de facto high-performance solutions
to graph problems. Then, since CPUs' clock speed increases slowed down,
multicore technology has become ubiquitous.
Recently, hardware accelerators such as GPUs (graphics processing units) and FPGAs
(field programmable gate arrays) have emerged to serve different parallelization
needs, and more are coming. The current computation environment is heterogeneous: it
consists of multicore servers with hardware accelerators. Such an environment
increases the importance of designing
flexible graph kernels that can run well on various platforms.

Systems researchers have proposed many graph processing frameworks to simplify
parallel graph algorithm design. Those systems
mostly adopts one of the
vertex-centric~\cite{Malewicz10-SIGMOD, Low12-VLDB,Kyrola12-OSDI,salihoglu13-ICSSDM,Shun13-PPOPP,Zhong13-TPDS,Gharaibeh13-ARXIV,Fu14-GRADES,Khorasani14-HPDC,Zhu15-ATC,Han15-VLDB,Sengupta15-SC,Sundaram15-VLDB,Kim16-SIGMOD,Wang16-SIGPLAN,Jia17-VLDB,Maass17-EUROSYS,Ma17-ATC},
edge-centric~\cite{Roy13-SOSP,Zhong13-TPDS,Sengupta15-SC,Wang16-SIGPLAN,Maass17-EUROSYS},
block-centric~\cite{Tian13-VLDB,Yan14-VLDB,Simmhan14-EuroPar,Hong17-PACT} or
linear-algebra-based~\cite{Kang09-ICDM,Kang11-KDD,Ghoting11-ICDE,Zhang16-IPDPSW,Kepner16-HPEC}
programming abstractions.
For a detailed discussion, we encourage the reader to see Section~\ref{sec:related}.
Most of the existing frameworks are designed to run
solely on one type of architecture, and they do not support heterogenous
execution. There exist a few shared-memory heterogeneous
systems~\cite{Gharaibeh13-ARXIV,Zhang15-JSC}. Those systems adopts vertex-centric
APIs.
Best of our knowledge \pgabb is
the first block-based heterogeneous graph processing framework.

A graph application may require running a variety of graph algorithms
in an analysis pipeline.
For instance, we may first run a
connected-components algorithm to find the largest component and extract it.
We may next run a BFS (breadth-first search) algorithm to re-order the
vertices, then a triangle counting algorithm to measure clustering coefficients
or more complicated features. All of the implementations in the
pipeline should perform well to be able to achieve good overall performance.
Vertex-centric and edge-centric programming abstractions are very restrictive and
restrictive programming models are not good for expressing a variety of
graph algorithms efficiently, and in general require complex
workarounds~\cite{Salihoglu14-GRADES}.

{\bf Key insights and contributions:}
In this work, we propose a novel graph processing framework for modern
shared-memory heterogeneous machines. Our framework has three design goals:
First, the framework can execute kernel operations of a graph computation on
different
architectures and combine the results. Second, the framework addresses major
parallel graph algorithm design challenges in the background.  Third, the
programming model is expressive to implement a diverse set of graph algorithms
that achieve competitive performance.

We claim that {\em a block-based programming model} is suitable to design
architecture-agnostic and efficient parallel graph algorithms
and we propose \pgabb (Parallel Graph Algorithms by Blocks) as our solution.
The literature includes a few block-based systems:
Giraph++~\cite{Tian13-VLDB}, Blogel~\cite{Yan14-VLDB}, and
GoFFish~\cite{Simmhan14-EuroPar} are three popular distributed graph
processing frameworks. Those systems use a connectivity-based partitioner to
reduce communication. They extend vertex-centric APIs: vertices in the same
block can access each other's information without communication.
MultiGraph~\cite{Hong17-PACT}, a graph processing
framework for GPUs, generates dense, light, and sparse edge blocks. It
uses a different data representation for each type of block.
This method allows MultiGraph to implement block-specific kernels.
\pgabb differs from those systems in the following ways:
First, the \pgabb user expresses the algorithm using a functor that
operates on an ordered list of blocks (\emph{block-list}),
not on a single block.
Second, \pgabb targets heterogeneous execution environments and does not
solely depend on a single architecture. Third, \pgabb uses
spatial partitioning.

In a block-based execution model, one might need to access data from multiple
blocks. Pattern-mining algorithms are good examples. For instance, the
triangle counting problem seeks the count of mutually-connected sets of
three vertices
(and their associated three edges). Edges of triangles may appear in three different blocks.
To process a block, we need two other blocks. There are two common
approaches for solving this issue: adding a communication step where
each worker requests data from other workers, or breaking the
decentralized computation model and making the graph (i.e., blocks) visible to
everyone. The former solution brings additional synchronization and message
passing costs. Many shared-memory systems~\cite{Shun13-PPOPP,Wang16-SIGPLAN}
applied the latter and achieved good performance. However, this approach is
not efficient for heterogeneous execution due to the limited memory of
hardware accelerators: they cannot process graphs larger than the device memory
size. To solve those problems, we introduce the block-list concept and build
our framework around it. In brief, a block-list is an ordered list of
references to blocks (i.e., sub-graphs) of a graph. The PGAbB user expresses
an algorithm using a functor that operates on a block-list
(see Subsection~\ref{sec:algs}).

Modeling the computation using block-lists brings several advantages.
First, we can process graphs larger than the device memory.
\pgabb only needs the blocks of a single block-list to process a task.
Second, we can localize the computation. Third, we can categorize graph
algorithms using three block-list generation styles that we specify and name:
\emph{activation-based}, \emph{single
block bulk-synchronous}, and \emph{multi-block pattern-based} (see
Section~\ref{ssec:pmodel}). Fourth, we can bound the required data movement to
the maximal size of a block-list.

The \pgabb user can run any computation on the host (CPU only), on the
device (GPU only), or both (collaborative).
In the collaborative setting: \pgabb generates tasks and sorts them based on
their
workload estimations. The goal is to detect bottleneck tasks. \pgabb assigns
bottleneck tasks to the device and the others to the host.
\pgabb uses streams and asynchronous calls to overlap the data copy time with
the computation.

\pgabb's programming model does not depend on a partitioning scheme.
However,  we encourage the usage of conformal two-dimensional spatial
partitioning, especially in heterogeneous settings. In data-driven
computations, conformal partitioning bounds the number of blocks that we
should fetch, makes reasoning about graph algorithms easier, and can localize
graph computation.

The contributions of our paper are as follows:

\begin{itemize}
\item We propose a novel block-based graph processing model directly motivated by
heterogeneous architectures.

\item We show that \pgabb can process graphs that state-of-the
art GPU based systems cannot.

\item We provide a classification method to categorize graph algorithms.

\item We analyze \pgabb's and prior arts' performances on $44$ real-world and
synthetic graphs against six state-of-the-art libraries/frameworks.
\end{itemize}

{\bf Experimental methodology and artifact availability.}
We have implemented five different graph algorithms on \pgabb and compared with
the implementations of these five algorithms on six state-of-the-art
libraries/framework. GAPBS~\cite{beamer15-gap}, a
hand-optimized lightweight graph library; three
CPU-based graph processing frameworks: Galois~\cite{kulkarni07-PLDI},
Ligra~\cite{Shun13-PPOPP}, LAGraph~\cite{davis19-TOMS}; and two
GPU-based graph processing frameworks:
Galois-GPU~\cite{burtscher12-IISWC}, and Gunrock~\cite{Wang16-SIGPLAN}.
All competitive frameworks (i.e., other than GAPBS) provides a
Vertex/Edge centric API. For parallelization, Galois,
Ligra and LAGraph use OpenMP, and Galois-GPU and Gunrock use CUDA.
Totem~\cite{Gharaibeh13-ARXIV} and
MultiGraph~\cite{Hong17-PACT} are designed for older GPU architectures therefore
were not able to include those systems in our experiments.
Best of our knowledge Blogel~\cite{Yan14-VLDB} outperforms
Giraph++~\cite{Tian13-VLDB}, and GoFFish~\cite{Simmhan14-EuroPar}.
However, on a single node Blogel was a magnitude of time
slower than \pgabb. Therefore we did not include
it in our experiments for fairness.

As we will present in more detain in Section~\ref{sec:exps}, we performed our
experiments on a heterogenous system with with $2$ Power~9 CPUs, and $2$
Nvidia Volta GPUs. And carried on experiments on 44 widely used benchmark graphs
that have edges from 100 millions to 2.1 billions.

After the double-blind review and our internal pre-release processes are
complete, we plan to release \pgabb on GitHub as open source under the BSD-3
license.

{\bf Limitations of the proposed approach.}
Our current \pgabb implementation is only for shared-memory heterogeneous
systems with CPUs and GPUs. \pgabb explicitly, and efficiently, manages the
data transfers between host and GPU memory. Adding different type accelerators
with their own local memory, and/or supporting distributed-memory execution will
necessitate modifications to \pgabb.

\section{Related Work}
\label{sec:related}

Programming abstractions of current frameworks can be broadly divided into
five categories: vertex-centric, edge-centric, block-centric, linear-algebra
based and domain-specific-language (DSL) based. Vertex-centric and
edge-centric programming abstractions allow flexibility to the framework
developers in system implementation by restricting the API~\cite{Dean04-OSDI}.
This trade-off is acceptable for some simple graph algorithms such as
PageRank~\cite{Page99-SI}, breadth-first search (BFS)~\cite{Beamer12-SC},
or label-propagation-based connected components. However, those API's
restrictive nature makes it hard to implement non-trivial graph algorithms
efficiently. One needs to break the decentralized computation model
to implement non-trivial graph
algorithms~\cite{kulkarni07-PLDI,Shun13-PPOPP,Salihoglu14-VLDB,davis19-TOMS}.

In recent years some linear-algebra-based graph processing
frameworks are proposed. The primary goal of these frameworks is to use
hand-optimized linear-algebra kernels at the back-end while users write their
algorithms in the language of linear algebra. This is a good idea however,
there are two drawbacks. First, one cannot use these frameworks without
linear-algebra knowledge. Second, in general, pure linear algebraic
formulations require more number of operations than graph-based solutions and
perform worse. Careful implementation is required to achieve a competitive
performance~\cite{Wolf17-HPEC} which requires algorithm-specific changes on
the linear-algebra kernels.

In the literature, there exists a few block-centric or block-based works.
Giraph++~\cite{Tian13-VLDB}, Blogel~\cite{Yan14-VLDB} and GoFFish~
\cite{Simmhan14-EuroPar} are three popular block-centric distributed graph
processing frameworks. The primary goal of those systems is to reduce
communication costs by using a connectivity-based partitioner to partition a
given graph among processors and extending vertex-centric API in a way that
vertices in the same block can access each other's information without any
communication. This approach has two problems. First, good connectivity-based
graph partitioners are computationally expensive. Second, those systems still
tend to have vertex-centric API restrictions but with some relaxation for the
vertices in the same block. MultiGraph~\cite{Hong17-PACT} proposes a different
block-based approach for GPU-based graph processing. MultiGraph generates
dense, light and sparse edge blocks using vertex order by degree. Besides,
Multigraph uses a different data representation for each type of block. This
method allows them to implement different kernels for each type of block for
maximal utilization. This approach has two problems. First, in this approach
reasoning on graph algorithms is hard due to the complex nature of blocks and
their representations. Second, MultiGraph uses a kind of jagged partitioning
which leads to higher communication costs on distributed and CPU/GPU
heterogeneous settings.


\section{Programming Model}
\label{ssec:pmodel}

\pgabb is is a header-only library. The \pgabb user implements their
code and functions in C++ using \pgabb's API.
The computation takes a set of disjoint blocks (i.e., sub-graphs) as input.
The user can define vertex, edge, and global attributes. There is no
pre-defined output format, the user can store output in an attribute.
The \pgabb user can provide six functors: two for doing the main computation,
one for block-list composition, two for controlling the iterative execution
process, and one for helping the scheduler.
Listing~\ref{template}
illustrates a high-level view of user-defined functors.

The user writes at least one of two kernels: a host kernel ($\mathcal{K}_H$),
and a device kernel ($\mathcal{K}_D$).  Both take as input a
block-list: an ordered list that contains references to blocks.
The user decides the number of blocks and the order of blocks in block-lists.
The host kernel ($\mathcal{K}_H$) runs on a CPU and the device kernel
($\mathcal{K}_D$) runs on a GPU. The user can access the data of all
blocks in the given block-list. \pgabb supports both CSR and COO
representations.

\begin{figure}
\begin{lstlisting}[ language=pgabb, escapeinside={(*}{*)}, caption=A
high-level view of a graph algorithm on \pgabb, label=template]
pgabb::start()
{
  /* Kernels: One of them has to be written */
  K_H = [](L_i)->void{}; // Host kernel
  K_D = [](L_i)->void{}; // Device kernel
  /* Block-lists: One of them has to be written */
  P_G = [](L_i)->bool{}; // Generic
  P_C = []()->L{}; // Custom
  /* Iterative execution */
  I_B = []()->void{};
  I_A = []()->bool{}; // Compulsory
  /* Workload estimation */
  E = [](L_i)->double{};
}
pgabb::stop();
\end{lstlisting}
\end{figure}

The \pgabb user can compose block-lists using two
different approaches. First, the user can generate all block-lists and provide
them to \pgabb with the \patterncus functor. Second, the user can
provide the \patterngen functor. That functor takes a block-list as input
and returns {\tt true} if the given block-list is a member of the computation.
In the latter, \pgabb checks all possible block-list combinations of the given
size and only keeps block-lists that \patterngen returns {\tt true}.

To support iterative computations the \pgabb user can use two functors:
\pre and \post. \pgabb executes the \pre functor before starting the
computation (\hostkernel, \devkernel). After the computation,
\pgabb executes the \post functor. The \post functor defines the termination
condition. \pgabb iterates while \post functor returns {\tt true}.

To customize the scheduling, the \pgabb user can provide an optional
estimation functor
\est. \est returns a weight for a given block-list.

In brief, designing an algorithm using \pgabb's programming model involves
four stages: block-list composition (\patterngen or \patterncus), attribute
assignment (vertex, edge and/or block), execution handling (\pre, \post, and
\est), and implementation of the kernel (\hostkernel and \devkernel).

\subsection{PGAbB Preliminaries}
\label{ssec:prelim}

A graph $G=(V, E)$, consists of a set of vertices $V$ and a set of
edges $E$. An edge $e$ is referred to as $e = (u, v) \in E$ where $u, v \in V$.
\pgabb assumes that the given graph, $G$, is divided into sub-graphs that we
call ``blocks.'' Let $B_i = (V_i=(S_i \cup D_i), E_i)$ denote a block.
We specify a block as three subsets: a subset of \emph{source
vertices} $S_i\subseteq V$, a subset of \emph{destination
vertices} $D_i \subseteq V$, and a subset of edges $E_i \subseteq E$.
For each edge $(u,v)\in E_i$, we have $u\in S_i$ and $v \in D_i$. We define
$B=\{B_0, B_1, \dots, B_k\}$ as the ground set of all blocks.
\pgabb does not replicate edges in different blocks. Therefore, all blocks
are disjoint and $B \equiv G$.

The \pgabb user can define vertex attributes, edge attributes and a global
attribute. Let $A_V$ denote the vertex attributes where $A_V(S_i)$ and $A_V
(D_i)$ represent source and destination vertex attributes of the block $B_i$,
respectively. Let $A_E$ denote the edge attributes where $A_E(E_i)$ represents
the edge attributes of the block $B_i$, and $A_G$ represents the global
attribute. Let $A_G$ denote the global attributes.

A block-list (e.g., $L_i=\langle B_j, B_k, B_l\rangle $), is an ordered list
of blocks and the user decides the size and the order of blocks in the
block-list. We denote $L=\{L_0, \dots, L_t\}$ as the set of all block-lists for
a graph algorithm. \pgabb models the computation using block-lists.

The computation of a kernel operation of a graph algorithm takes a
block-list $L_i$ as
input.
The \pgabb user provides the \hostkernel functor for CPUs and
\devkernel for GPUs. Both functors take a block-list as input. A kernel and a
block-list define a task, $T_i$, such that;
$T_i=$\hostkernel($L_i$) or $T_i=$\devkernel($L_i$). $T$ represents the set of
all tasks.

\subsection{Categorization of graph algorithms in \pgabb}
\label{sec:algs}

\begin{figure}[ht]
  \centering
  \includegraphics[width=.45\linewidth]{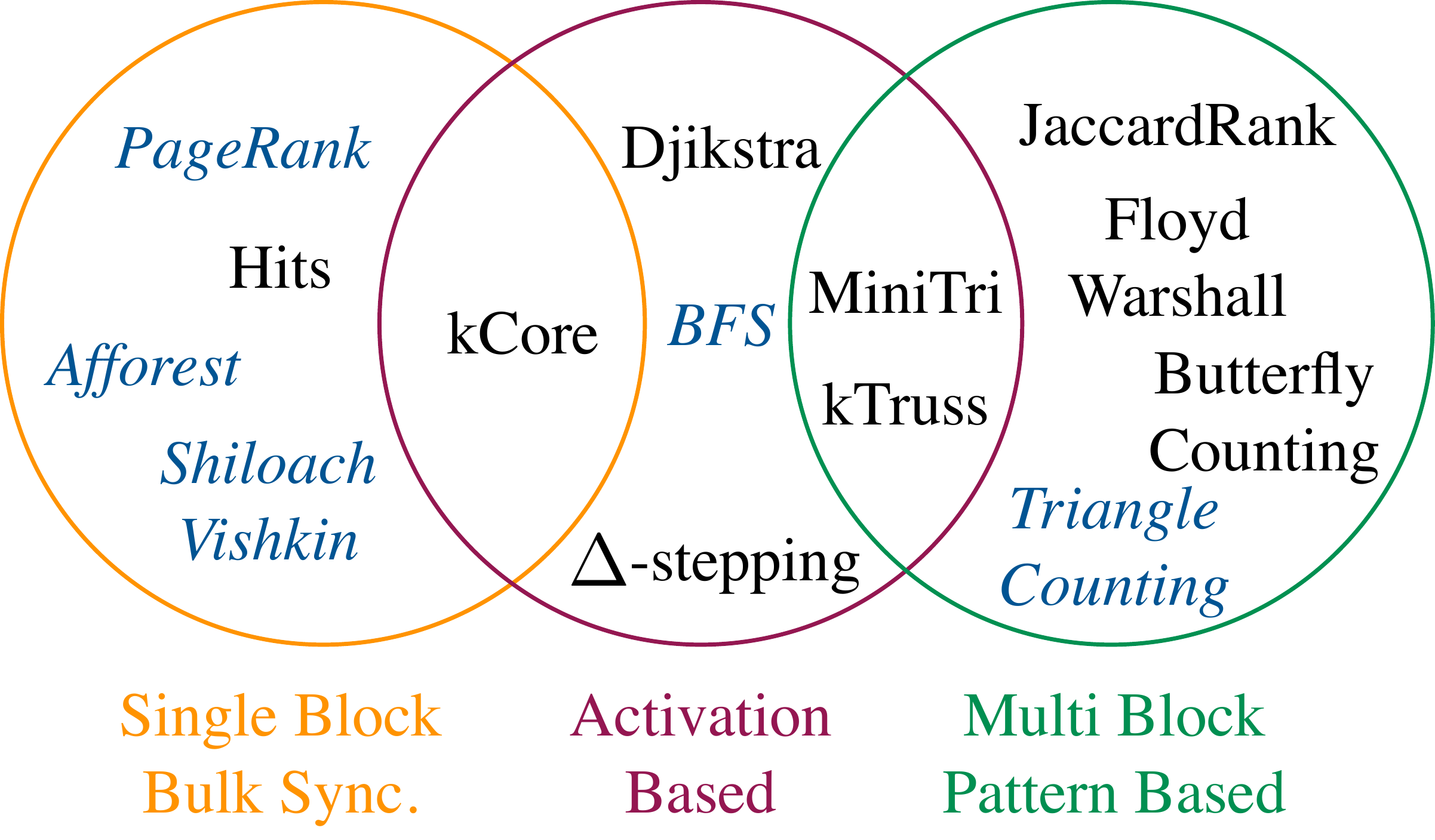}
  \caption{A classification of some graph algorithms in \pgabb.
            We will cover the ones written in italic font.}
  \label{fig:algsep}
\end{figure}

Parallel algorithm design for irregular graphs is a challenging
research problem. Therefore, in the literature there exist many
algorithms to solve the same graph problem. In general, there
is no clear winner because both the structure of the graph and the nature of
the architecture impact performance. To accommodate these variables,
API permissiveness is important. On the other hand, more restrictive APIs
can help system developers~\cite{Dean04-OSDI}. To become more
permissive \pgabb supports three execution modes: {\em single block bulk
synchronous execution}, {\em activation-based execution}, and {\em multi-block
pattern-based execution}.
Fig.~\ref{fig:algsep} illustrates some of the
popular graph algorithms and which \pgabb execution mode is a good fit for
that algorithm.

\noindent {\bf Activation-based execution:} Many graph computations are
iterative and data-driven. In this execution model, the next iteration's
computation (i.e., active
vertices/edges) depends on the vertices visited in the current iteration.
Therefore, only some portion of the graph contributes to the computation.
Traversal (i.e., BFS, Dijkstra~\cite{Dijkstra59-NM}) and peeling-based graph
algorithms (i.e., kTruss~\cite{Cohen08-NSA}) are good examples for that
execution mode.

\noindent {\bf Single block bulk synchronous execution:} Another common
execution pattern among different graph algorithms is visiting all of the
edges in an iterative fashion. PageRank~\cite{Page99-SI},
HITS~\cite{Kleinberg99-CSUR}, Shiloach-Vishkin~\cite{Shiloach82-JA} algorithms
are good examples for this execution mode. The computation visits all of the
edges in each iteration. Therefore, in this execution mode composing block
lists from single blocks is sufficient to obtain efficient parallelism.

\noindent {\bf Multi-block pattern-based execution:} Non-trivial
graph algorithms that have more complicated computation structure may require
to access different parts of the graph. Therefore we need to compose block
lists using multiple blocks depending on the pattern of the computation
structure. Triangle counting~\cite{Latapy08-TCS}, k-clique finding, and
butterfly counting algorithms are good examples for this execution mode.

Note that those execution modes are not strict and a graph algorithm may use
two execution modes.

\subsection{Parallel dispatches and basic API}
\label{ssec:api}

\pgabb provides parallel for loop, and parallel reduction primitives for host
and device. The \pgabb user can define the parallel body of those primitives
using functors or lambdas.
Parallel for loop primitives; \kwverb|for_host| and \kwverb|for_dev|,
take the size of
loop, and functor (or lambda function) written by the user. Parallel reduction
primitives;
\kwverb|reduce_host| and \kwverb|reduce_dev|,
take an additional variable to represent reduction variable.

The \pgabb user can access source and destination vertices of the block $B_i$
using \emphverb|Vertices|~($S_i$ | $D_i$). \emphverb|Edges|~($E_i$, $u$)
returns neighbors of the  vertex $u\in S_i$ in the block $B_i$.
\emphverb|Attribute|~($A_V$($S_i$ | $D_i$)) returns attributes of source or
destination vertices of $B_i$.
\emphverb|Attribute|~($A_E$($E_i$)) returns edge attributes of $B_i$, and the
global attributes can be accessed using \emphverb|Attribute|~($A_G$).

\pgabb can do all read and write operations atomically. In this case the
user does not need to worry about race conditions. However, atomic operations
are expensive.
To avoid this overhead, \pgabb provides atomic functions to users who have
basic parallel programming knowledge: \emphverb|Add|~(a, b);
atomically adds $b$ to $a$. \emphverb|CAS|~(a, b, c), compare-and-swap, takes
three arguments; a memory location ($a$), an old value ($b$) and a new value
($c$). If the value stored at $a$ is equal to $b$ it atomically swaps $c$ with
$a$ and returns {\tt true}, and otherwise it does not update $a$ and returns
{\tt false}.

\begin{figure}[ht]
\begin{lstlisting}[language=pgabb, escapeinside={(*}{*)}, caption=Shiloach-Vishkin device kernel on \pgabb., label=sv-dev]
auto K_D = [](BlockList L_i){
  auto B_i = L_i(0);
  // Global attribute
  auto [C, H] = Attribute(A_G);
  if (pgabb::Iteration % 2 == 0){
    reduce_dev(Vertices(S_i).size(),[](auto u){(*\label{sv-l1}*)
      int h = 0; // local number of hooks
      for (auto v : Edges(E_i, u)){
        auto r1 = max(C(u), C(v)); // max root
        auto r2 = min(C(u), C(v)); // min root
        if (r1 == r2) continue;
        if (r1 == C(r1)){
          C(r1) = r2;
          ++h;
        }
      }
      H = H + h;
    }, H);
  } else{
    auto [u, v] = GetInterval(i, |C|)
    for_dev(v-u, [](auto k){ (*\label{sv-l2}*)
        while (C(u+k) != C(C(u+k)){
            C(u+k) = C(C(u+k));
        }
    });
  }
};
\end{lstlisting}
\end{figure}

\subsection{An example: Shiloach-Vishkin's algorithm.}
\label{ssec:sv-ex}

In this section we are going to explain how the \pgabb user can implement
Shiloach-Vishkin's~\cite{Shiloach82-JA} algorithm to solve weakly connected
components problem. This is an iterative algorithm consists of two steps.
In the first step, the algorithm iterates over edges and combines vertices
into trees: for each edge, if roots of source and destination vertices are
different, then the algorithm tries to hook the greater root to the smaller
root. In the second step, the algorithm links each vertex to the root of its
tree using pointer jumping. Algorithm stops when there is not any hooking
operation performed.

\noindent {\bf Block-list composition:} The \pgabb user designs the
computation using block-lists. Therefore, the first step is the
decision of block-list composition: Number of blocks per block-list, and their
order. Shiloach-Vishkin's algorithm iterates all edges during the hooking
step. Therefore, we can set block-list size to one, so each block-list has
a block. Providing \patterngen functor that returns {\tt true} is sufficient.

\noindent {\bf Attribute assignment:} Note that within a block the user can
access source vertex, destination vertex, and edge attributes. However,
pointer jumping may jump outside vertices. Therefore
we are going to assign and use global attributes: An array of size $|V|$; $C$,
to store parent ids of each vertex. A variable, $H$, to store number of
hook updates.

\noindent {\bf Execution handling:} Shiloach-Vishkin algorithm is iterative
and each iteration consists of hooking and linking steps. In our design,
during the even iterations we do the hooking and during the odd iterations we
do the linking:
Hook $\rightarrow$ Link $\rightarrow \dots \rightarrow$ Hook $\rightarrow$ Link.
In \pre functor, before each hooking iteration we should reset $H$ to $0$.
In \post functor, we should check if a hooking operation is done during the
hooking iteration if so return {\tt true} otherwise return {\tt false}.
We are going to use number of edges in a block-list for estimating the
workload: \est functor returns the number of edges in a block-list.

\noindent {\bf Kernel development:} Listing~\ref{sv-dev} illustrates an
implementation of \devkernel. In our implementation, first, we get the block
reference, $B_i$, from the block-list, $L_i$. Note that, each block-list has
one block. Then we fetch block attributes, $C$ and $H$. After, we check
whether we are in an odd or even iteration. If it is even, then we iterate all
edges in the block and do hooking operations. Here we store local number of
hook operations, $h$, and at the end add it atomically to $H$. If we are in an
odd iteration then we do the linking operations. Linking operations depends on
the $C$ array. Therefore, first we should divide the array $C$ equally among
different block-lists for parallel processing. The \pgabb user can use
{\tt GetInterval(id, size)} function to get a unique interval. Such an
approach becomes highly useful when we partition the graph in two-dimension.
\hostkernel differs only in outer loop execution; line~\ref{sv-l1} and
line~\ref{sv-l2} in Listing~\ref{sv-dev}. \devkernel uses parallel reduction and for loop dispatches
to leverage from the massive parallelization capabilities of GPUs. In
\hostkernel we use regular for loops.

\begin{figure}
  \begin{lstlisting}[language=pgabb, escapeinside={(*}{*)}, caption=BFS device kernel on \pgabb., label=bfs-dev]
auto K_D = [](BlockList L_i){
  auto B_i = L_i(0);
  auto [Q, n_q] = Attribute(A_G); // attributes
  auto P = Attribute(A_V(S_i));   // BFS Tree
  size_t n = 0;
  reduce_dev (S_i.size(), [](auto u, auto n){ (*\label{bfs-l1}*)
    for (auto v : Edges(E_i, u)){
      if (Q(D_i)[v] && CASS(P[u], -1, v)){
        Q(S_i).push(u);
        ++n;
        break;
      }
    }
  }, n_q);
};
  \end{lstlisting}
\end{figure}

\subsection{An example: direction-optimized breadth-first search (BFS) algorithm}
\label{ssec:bfs}

The BFS problem is to create a tree starting from a source vertex such that
each level represents the distance between the source vertex, and every other
vertex. The BFS algorithm explores all nodes at the present depth prior to
moving on to the nodes at the next depth level. In each level, we insert
unvisited neighbors of vertices in the frontier queue, to the next queue.
In undirected graphs, one can perform this operation in two ways: Top-down:
for all vertices in the current queue adding unvisited neighbors to
the next queue. Bottom-up: visiting all vertices and adding them to the
next-level queue if one of their neighbors is in the current queue.
Beamer et al.~\cite{Beamer12-SC} proposed to use a direction
optimization for traversing less number of edges when the frontier
queue becomes large. We are going to implement a version of this algorithm.

\begin{figure}[ht]
\begin{lstlisting}[language=pgabb, escapeinside={(*}{*)}, caption=BFS host kernel on \pgabb., label=bfs-host]
auto K_H = [](BlockList L_i){
  auto B_i = L_i[0];
  auto [Q, n_q] = Attribute(A_G); // attributes
  auto P = Attribute(A_V(D_i));   // BFS Tree
  size_t n = 0;
  for (auto u : Q(S_i)){
    for (auto v : Edges(E_i, u)){
      if (CASS(P[v], -1, u)){
        Q(S_i).push(v);
        ++n;
      }
    }
  }
  Add(n_q, n);
};
\end{lstlisting}
\end{figure}

\noindent {\bf Attribute assignment:} We are going to use vertex attributes to
store the parent of each vertex. We are going to use a level-based
queue data structure, $Q$, as the global attribute. Depending on the partitioning
scheme, a block might use its own queue only. Therefore, assume that, $Q(S_i)$ and
$Q(D_i)$ represent queues of source and destination vertices of the
block, $B_i$, respectively. In addition we are going to use a variable,
$n_q$, to count the number of vertices that we inserted in an iteration.

\noindent {\bf Block-list composition:} Each block composes
a block-list. We are going to use activation-based execution model.
Therefore, in each iteration we are going to create block-lists from the
blocks whose source or destination vertex queues are not empty.

\noindent {\bf Execution handling:} We are not going to use the \pre functor
for this algorithm, and the \post functor is going to finalize the iterations
when we did not push any vertices to $Q$.

\noindent {\bf Kernel development:} By their nature GPUs are more suitable to
run bottom-up BFS algorithm while CPUs are better at top-down BFS algorithm.
Therefore, in our design we are going to implement top-down BFS in \hostkernel
and bottom-up BFS in \devkernel. In \hostkernel, after fetching the block
information and corresponding queue data-structure we are going to visit
neighbors of each vertex in the frontier and push them to our queue if they
are not visited before. This process continues until there is not any push
operation in an iteration. Listing~\ref{bfs-host} illustrates this algorithm.
In \devkernel, we implement bottom-up BFS algorithm. In this kernel,
after fetching the block information and required attributes, we visit all of
the vertices. If we have not visited a vertex and one of its neighbors appears
in the frontier then we insert that vertex to our queue and stop visiting the
other neighbors. Again we keep track the number of push operation to
understand if we have finished execution or not. Listing~\ref{bfs-dev}
illustrates this kernel.

\subsection{An example: triangle counting algorithm}
\label{ssec:tc}

\begin{figure}
\begin{lstlisting}[language=pgabb, escapeinside={(*}{*)}, caption=Triangle Counting device kernel on \pgabb, label=tc-dev]
auto K_D = [](BlockList L_i){
  auto B_k = L_i(0);
  auto B_l = L_i(1);
  auto B_m = L_i(2);
  size_t n_t = Attribute(A_G);
  reduce_dev (S_k.size(), [](auto u, auto n_t){(*\label{tc-l}*)
    for (auto v : Edges(E_k, u)) {
      // Partial neighborlist of u
      auto N_u = Edges(E_l, u);
      // Partial neighborlist of v
      auto N_v = Edges(E_m, v);
      // Intersect: number of common vertices
      n_t += Intersect(N_u, N_v);
    }
  }, n_t);
};
\end{lstlisting}
\end{figure}
  
The triangle counting problem is to find the number of mutually connected
sets of three vertices in an undirected graph. Common approach to solve this problem
is to compute common neighbors between source and destination vertices of each
edge. In this work we are going to implement a two-dimensional triangle counting
algorithm~\cite{Yasar21-TPDS}.

\noindent {\bf Block-list composition:} Each block-list is going to have three
blocks. For each edge in the first block, the second block is going to contain
source vertex partial-adjacency list and the third block is going to contain
destination vertex partial-adjacency list: $L_i=(B_k, B_l, B_m)$ such that
$S_l = D_k$ and $S_m = D_l$.

\noindent {\bf Attribute assignment:} We are going to use a variable, $n_t$, as
the global attribute to store the number of triangles found.

\noindent {\bf Execution handling:} This algorithm does not require iterative
execution. Therefore, we are not going to use \pre and \post functors.

\noindent {\bf Kernel development:} The goal is counting the number of
triangles that appear in three blocks of a block-list, $L_i=(B_k, B_l, B_m)$.
For each edge, $(u,v)$ in $E_k$, this operation can be done by counting the
number of common neighbors between the partial-neighbor list of $u$ in $B_l$
and the partial neighbor list of $v$ in $B_m$. Similar to regular triangle
counting algorithms this can be computed using list or hashmap-based
intersection algorithms~\cite{Latapy08-TCS}. Listing~\ref{tc-dev}
illustrates \devkernel. We use parallel reduction to achieve parallelization
within a block-list on GPUs. \hostkernel uses regular for loop
and differs only in line~\ref{tc-l} of Listing~\ref{tc-dev}.

\begin{figure*}[ht]
  \centering
  \includegraphics[width=.7\linewidth]{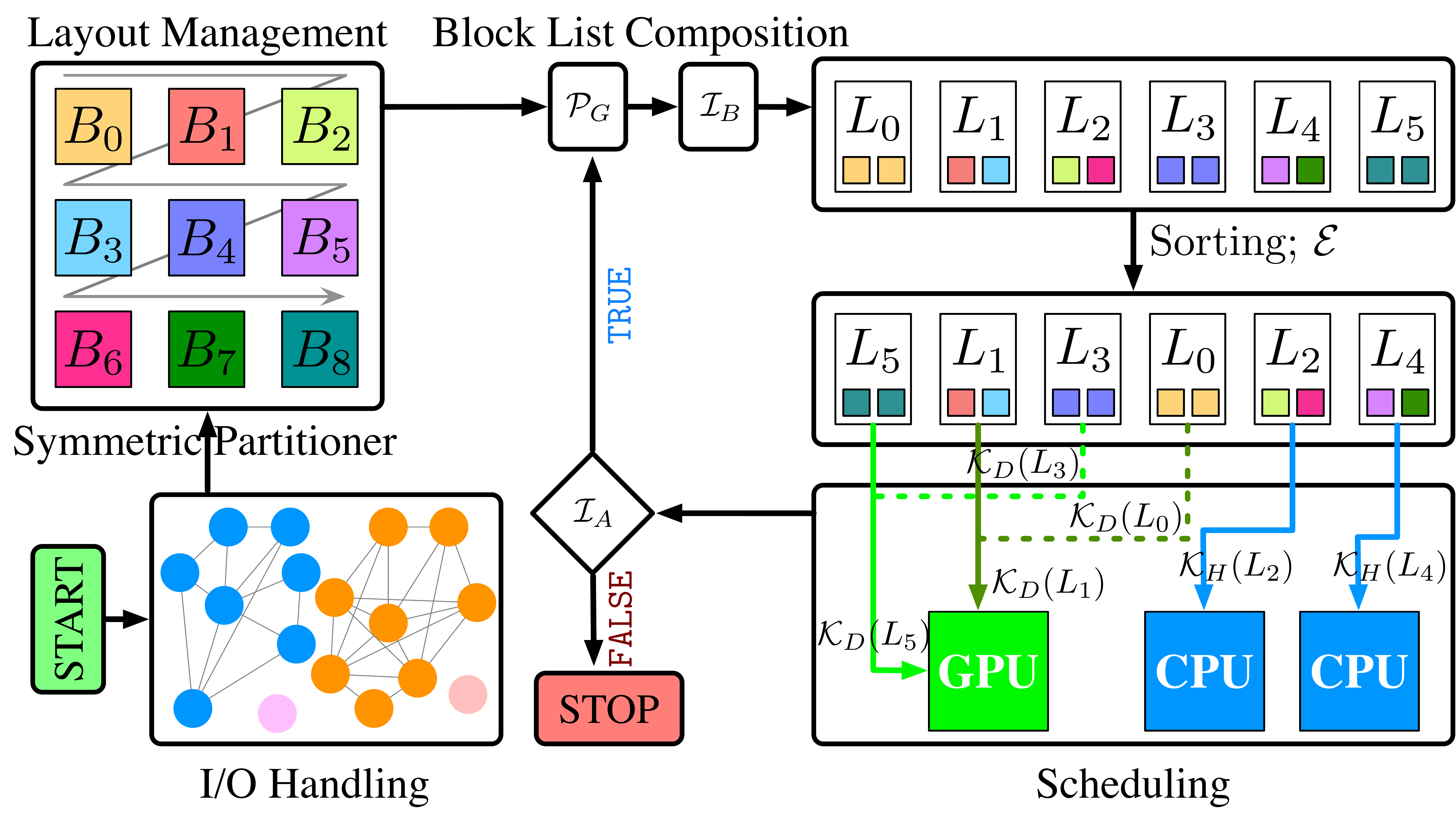}
  \caption{High-level execution flow.}
  \label{fig:exec}
\end{figure*}

\section{Design of the  PGAbB Framework}
\label{sec:framework}

We target single node heterogeneous environments: multicore CPUs and a
GPU. \pgabb consists of four components; I/O Handler, Layout Manager \&
Partitioner, Scheduler, and a user API.
As backend, we use Kokkos~\cite{Edwards14-JPDC}; OpenMP on the host machine
and CUDA on the device.

\subsection{Execution flow}

First, \pgabb reads the graph from disk and
partitions it into blocks. The user can use one of the
existing partitioners or provide a custom partitioner. After the block
composition, \pgabb assigns an id to each block and orders them.

Generation of block-lists is the third step. Using \patterncus or \patterngen,
\pgabb generates block-lists, assigns them a weight using \est functor and
sorts them in decreasing order. The goal is distinguishing bottleneck
tasks to be able to assign them to GPU. \pgabb aims to assign lighter tasks
to CPUs.

\pgabb parallelizes a computation using tasks and utilizes CPUs and GPU
streams. A CPU thread executes a single task. A GPU stream uses multiple
threads to execute a task.
After the ordering, \pgabb initialized the scheduler and executes \pre
functor. Then, it assigns heavy tasks to GPU streams, and light ones to CPUs.
When all tasks are processed, \pgabb executes \post functor. If \post functor
returns {\tt false} \pgabb stops, otherwise starts a new iteration.
Fig.~\ref{fig:exec} illustrates high-level view of this execution flow.

\subsection{I/O Handler}
\label{ssec:io}

Most of the graph kernels are computationally lightweight, and disk I/O can
become the bottleneck. Therefore, almost all of the graph processing
frameworks use custom binary format to store the graph. \pgabb also
uses a custom binary format to make reading process faster. In addition,
to be able to read ASCII formatted graph inputs in the most efficient way,
\pgabb adopts and extends PIGO~\cite{Gabert21-GrAPL}, a library for reading
and writing ASCII formatted graph files in parallel.

\subsection{Partitioner and Layout Manager}
\label{ssec:parti}

\pgabb does not dictate any partitioning scheme to form blocks. However, based
on our experiences, we strongly encourage the use of a symmetric
two-dimensional partitioner~\cite{Yasar20-ARXIV-SARMA}
in hybrid settings for three reasons: First,
gathering information and scattering computation results become
straightforward. Gathering information from the in-edges of the vertices, computing on those
and then scattering the result to all the processors that has the outer-edges
is a common communication pattern in graph analysis, such as
BFS~\cite{awerbuch1985distributed}, and PageRank~\cite{sarma2015fast}.
Second, this type of partitioning becomes highly useful to reason about graph
algorithms: Each block is a sub-graphs where diagonal blocks are the owners of
the vertex meta-data and any other block represents the edges between two
sub-graphs.
Third, we can define a partitioning as conformal when connecting row and
column lengths of different tiles match. Conformal partitioning is very
crucial for many graph applications that access neighbors of neighbors or
tiled matrix-matrix multiplication based operations. Because, for those kind
of applications if one does not use a conformal partitioning for inputs and
outputs, converting outputs of the previous iteration to inputs of next
iteration would require additional communication. By its nature, symmetric
rectilinear partitioning is a conformal partitioning and it is suitable for
those applications.

Besides, one dimensional or vertex partitioning is very useful for CPU
only execution because it increases the locality of threads. Therefore, \pgabb
also provides a one-dimensional optimal partitioning algorithm. The \pgabb
user can also override \pgabb partitioner class.

\subsubsection{Block layout:} \pgabb assigns an integer ID for each
block and align them from the lowest to the largest. By default, \pgabb uses
row-major order to align blocks (see Fig.~\ref{fig:exec}). The \pgabb user
can override the partitioner and assign IDs using any space-filling curve
mapping or another method.

\subsubsection{Block storage data-structures:} Each block represents a
subgraph. By default \pgabb rearranges vertex ids within a block and uses
Compressed Sparse Row (CSR) data structure to store sub-graphs. In addition,
\pgabb also supports Coordinate List (COO), and Compact Coordinate List (CCOO)
data structures. Depending on the algorithmic need the \pgabb user can
initiate those data structures.

\subsection{Scheduler}
\label{ssec:sched}

Scheduler is responsible for assigning tasks to the GPU and CPUs.
\pgabb stores the graph in the host memory. Therefore, the scheduler
copies blocks in the BlockList of a task to the GPU memory after
assigning that task to the GPU. We use asynchronous memory copies
from host to device (device to host) to perform those operations.
When availabe GPU memory is under a threshold, scheduler waits
GPU tasks' termination, then clears the GPU memory before assigning
new tasks.
While designing \pgabb, we also experimented using UVM (Unified Virtual Memory)
but the performance was $\approx 5\times$ slower. Therefore,
we do not use UVM or present UVM-based results in this paper.

Our scheduler aims to assign computationally heavy tasks to the GPU and
lighter tasks to CPUs. The goal is leveraging the massive parallelism
available on GPUs.
\pgabb orders tasks based on their estimations and starts assigning heavier
tasks to GPUs, and lighter ones to CPUs. For sorting, \pgabb estimates
the workload of a task, $T_i$, using \est functor if defined, otherwise \pgabb
uses the total number of edges within a block-list as the weight of a task.
To decrease the initial synchronization cost of the GPU and guarantee the
assignment of heavier tasks to the GPUs, a cut-off can be predefined. CPUs do
not go past the cut-off.

We use CUDA streams to execute several tasks on the GPU simultaneously.
Increasing the number of CUDA streams, initially helps to improve the
performance and then starts hurting due to scheduling overhead. We
experimentally selected four as the number of CUDA streams. We create four
CUDA streams, then one of the CPU threads is assigned
to a stream. That CPU thread is responsible for waiting on that stream,
synchronizing the stream, sending a task to a device through that stream, and
gathering information from a device through that stream. All of these
operations use asynchronous function calls. When we create streams and assign
a thread for each of them, GPUs and CPUs compete for tasks and get a new one
from the queue when they finish executing a task. Then a stream thread can
overlap copying the blocks of the next assignment with the computation.

\pgabb's ultimate goal is to simultaneously leverage both CPUs and GPU. Such
an approach is crucial for achieving three goals: maximizing memory
utilization, processing graphs that cannot fit into GPU memory, and
overlapping the data-copy time with the computation. However, in some
lightweight graph kernels such an approach might not be suitable. Therefore,
the \pgabb user can choose to run computation using only CPUs or the GPU. The
user can do this decision before starting an iteration using \pre functor.

\section{Experimental Evaluation}
\label{sec:exps}

We performed our experiments on a Power9 architecture with $2$ CPUs, and $2$
Volta GPUs. Each CPU has $16$ cores and each core has $4$ SMTs  (Simultaneous
Multi-Threading). The machine has $320$GB memory and each Volta GPU has $32$
GB memory. We measured data transfer rate between CPU and GPU as
$\approx11$GB/s with pageable memory and $\approx60$ GB/s with pinned memory.
We compiled codes using GNU compiler (g++) version $7.4$, CUDA
runtime version $10.0$ and OpenMP version $4.0$.

In our experiments, we ran each implementation ten times and we report the
median of them. We only report the execution time of the graph kernels.
In all systems, we excluded disk I/O and other pre-processing overheads.
Data transfer overheads between the host and the device are included
in \pgabb's execution time but for others it was considered as part of the
pre-processing time.

\begin{table*}[ht]
\small
\caption{Speedups over the GAPBS reference implementation.
Each cell represents a framework's speedup on a graph and a graph algorithm.
Heat map indicates where speedup is lower (RED $< 1.0$), equal
(WHITE $=1.0$) or higher (GREEN $> 1.0$).
\pgabb-GPU: GPU-only results. \pgabb: Results with CPU and GPU.}
\begin{center}
\begin{tabular}{| c | l | C{1.8cm} | C{1.8cm} | C{1.8cm} | C{1.8cm} | C{1.8cm} | C{1.8cm} | C{1.8cm} |}
\cline{3-9}
 \multicolumn{2}{c|}{} & \multicolumn{2}{c|}{\bf Social} & \multicolumn{1}{c|}{\bf Web} & \multicolumn{1}{c|}{\bf Gene} & \multicolumn{1}{c|}{\bf Road} & \multicolumn{2}{c|}{\bf Synthetic}  \\ \cline{3-9}
 \multicolumn{2}{c|}{} & \bf twitter7 & \bf Orkut & \bf sk-2005 & \bf kmer\_V1r & \bf eu\_osm & \bf myciel19 & \bf kron21  \\ \hline
\hline
\parbox[t]{2mm}{\multirow{5}{*}{\rotatebox[origin=c]{90}{\bf Galois}}}
 & PR
     & \cellcolor{red!16}\bf0.83
     & \cellcolor{green!10}\bf1.01
     & \cellcolor{green!10}\bf1.01
     & \cellcolor{red!10}\bf0.89
     & \cellcolor{green!10}\bf1.03
     & \cellcolor{green!69}\bf6.96
     & \cellcolor{red!21}\bf0.78
 \\
 & SV/LP
     & \cellcolor{green!83}\bf8.40
     & \cellcolor{green!17}\bf1.71
     & \cellcolor{green!16}\bf1.68
     & \cellcolor{green!22}\bf2.29
     & \cellcolor{green!18}\bf1.81
     & \cellcolor{green!12}\bf1.25
     & \cellcolor{green!11}\bf1.12
 \\
 & CC
     & \cellcolor{red!16}\bf0.84
     & \cellcolor{green!15}\bf1.56
     & \cellcolor{red!2}\bf0.98
     & \cellcolor{red!36}\bf0.64
     & \cellcolor{red!35}\bf0.64
     & \cellcolor{green!29}\bf2.94
     & \cellcolor{red!18}\bf0.81
 \\
 & BFS
     & \cellcolor{red!74}\bf0.26
     & \cellcolor{red!40}\bf0.59
     & \cellcolor{red!53}\bf0.46
     & \cellcolor{red!66}\bf0.34
     & \cellcolor{green!21}\bf2.14
     & \cellcolor{red!61}\bf0.39
     & \cellcolor{red!81}\bf0.18
 \\
 & TC
     & \cellcolor{red!30}\bf0.69
     & \cellcolor{green!10}\bf1.06
     & \cellcolor{red!36}\bf0.63
     & \cellcolor{red!10}\bf0.90
     & \cellcolor{green!12}\bf1.21
     & \cellcolor{red!55}\bf0.44
     & \cellcolor{red!60}\bf0.40
 \\
\hline
\hline
\parbox[t]{2mm}{\multirow{5}{*}{\rotatebox[origin=c]{90}{\bf Ligra}}}
 & PR
     & \cellcolor{red!61}\bf0.39
     & \cellcolor{red!39}\bf0.60
     & \cellcolor{red!1}\bf0.99
     & \cellcolor{red!56}\bf0.43
     & \cellcolor{red!46}\bf0.53
     & \cellcolor{green!25}\bf2.59
     & \cellcolor{red!28}\bf0.72
 \\
 & SV/LP
     & \cellcolor{green!12}\bf1.24
     & \cellcolor{red!29}\bf0.70
     & \cellcolor{green!10}\bf1.05
     & \cellcolor{red!81}\bf0.18
     & \cellcolor{red!98}\bf0.02
     & \cellcolor{red!41}\bf0.58
     & \cellcolor{red!34}\bf0.66
 \\
 & CC
     & \cellcolor{red!97}\bf0.02
     & \cellcolor{red!95}\bf0.04
     & \cellcolor{red!100}\bf0.00
     & \cellcolor{red!98}\bf0.02
     & \cellcolor{red!99}\bf0.01
     & \cellcolor{red!97}\bf0.03
     & \cellcolor{red!98}\bf0.02
 \\
 & BFS
     & \cellcolor{red!38}\bf0.61
     & \cellcolor{red!33}\bf0.67
     & \cellcolor{red!7}\bf0.93
     & \cellcolor{red!31}\bf0.68
     & \cellcolor{red!83}\bf0.16
     & \cellcolor{green!13}\bf1.37
     & \cellcolor{red!18}\bf0.82
 \\
 & TC
     & \cellcolor{red!68}\bf0.31
     & \cellcolor{red!65}\bf0.35
     & \cellcolor{red!87}\bf0.12
     & \cellcolor{red!69}\bf0.30
     & \cellcolor{red!83}\bf0.17
     & \cellcolor{red!56}\bf0.43
     & \cellcolor{red!31}\bf0.69
 \\
\hline
\hline
\parbox[t]{2mm}{\multirow{5}{*}{\rotatebox[origin=c]{90}{\bf LAGraph}}}
 & PR
     & \cellcolor{red!24}\bf0.75
     & \cellcolor{red!1}\bf0.98
     & \cellcolor{red!39}\bf0.60
     & \cellcolor{red!25}\bf0.75
     & \cellcolor{red!34}\bf0.65
     & \cellcolor{green!32}\bf3.21
     & \cellcolor{red!28}\bf0.71
 \\
 & SV/LP
     & \cellcolor{green!100}\bf14.24
     & \cellcolor{green!16}\bf1.64
     & \cellcolor{red!11}\bf0.89
     & \cellcolor{red!70}\bf0.30
     & \cellcolor{red!87}\bf0.13
     & \cellcolor{green!76}\bf7.70
     & \cellcolor{red!8}\bf0.92
 \\
 & CC
     & \cellcolor{red!83}\bf0.17
     & \cellcolor{red!79}\bf0.21
     & \cellcolor{red!88}\bf0.12
     & \cellcolor{red!85}\bf0.14
     & \cellcolor{red!95}\bf0.05
     & \cellcolor{red!73}\bf0.27
     & \cellcolor{red!91}\bf0.09
 \\
 & BFS
     & \cellcolor{red!21}\bf0.79
     & \cellcolor{red!66}\bf0.33
     & \cellcolor{red!23}\bf0.77
     & \cellcolor{red!73}\bf0.27
     & \cellcolor{red!67}\bf0.33
     & \cellcolor{red!25}\bf0.75
     & \cellcolor{red!70}\bf0.30
 \\
 & TC
     & \cellcolor{red!61}\bf0.38
     & \cellcolor{red!12}\bf0.87
     & \cellcolor{red!34}\bf0.66
     & \cellcolor{red!71}\bf0.29
     & \cellcolor{red!84}\bf0.16
     & \cellcolor{red!48}\bf0.52
     & \cellcolor{red!63}\bf0.37
 \\
\hline
\hline
\parbox[t]{2mm}{\multirow{5}{*}{\rotatebox[origin=c]{90}{\bf Galois-GPU}}}
 & PR
     & \cellcolor{red!100}\bf0.00
     & \cellcolor{green!27}\bf2.72
     & \cellcolor{red!100}\bf0.00
     & \cellcolor{green!10}\bf1.01
     & \cellcolor{green!14}\bf1.49
     & \cellcolor{green!100}\bf12.12
     & \cellcolor{green!16}\bf1.62
 \\
 & SV/LP
     & \cellcolor{red!100}\bf0.00
     & \cellcolor{green!36}\bf3.67
     & \cellcolor{red!100}\bf0.00
     & \cellcolor{green!24}\bf2.43
     & \cellcolor{green!27}\bf2.71
     & \cellcolor{green!26}\bf2.65
     & \cellcolor{green!15}\bf1.57
 \\
 & CC
     & \cellcolor{red!100}\bf0.00
     & \cellcolor{red!53}\bf0.46
     & \cellcolor{red!100}\bf0.00
     & \cellcolor{green!11}\bf1.16
     & \cellcolor{red!1}\bf0.99
     & \cellcolor{red!90}\bf0.09
     & \cellcolor{red!85}\bf0.15
 \\
 & BFS
     & \cellcolor{red!100}\bf0.00
     & \cellcolor{red!100}\bf0.00
     & \cellcolor{red!100}\bf0.00
     & \cellcolor{red!100}\bf0.00
     & \cellcolor{red!100}\bf0.00
     & \cellcolor{red!100}\bf0.00
     & \cellcolor{red!100}\bf0.00
 \\
 & TC
     & \cellcolor{green!10}\bf1.03
     & \cellcolor{red!14}\bf0.85
     & \cellcolor{red!9}\bf0.90
     & \cellcolor{red!99}\bf0.00
     & \cellcolor{red!99}\bf0.00
     & \cellcolor{red!61}\bf0.38
     & \cellcolor{red!34}\bf0.65
 \\
\hline
\hline
\parbox[t]{2mm}{\multirow{5}{*}{\rotatebox[origin=c]{90}{\bf Gunrock}}}
 & PR
     & \cellcolor{red!100}\bf0.00
     & \cellcolor{green!12}\bf1.28
     & \cellcolor{red!100}\bf0.00
     & \cellcolor{green!14}\bf1.44
     & \cellcolor{green!13}\bf1.34
     & \cellcolor{green!54}\bf5.42
     & \cellcolor{red!2}\bf0.97
 \\
 & SV/LP
     & \cellcolor{red!100}\bf0.00
     & \cellcolor{green!18}\bf1.88
     & \cellcolor{red!100}\bf0.00
     & \cellcolor{green!31}\bf3.18
     & \cellcolor{green!12}\bf1.22
     & \cellcolor{green!38}\bf3.90
     & \cellcolor{red!2}\bf0.97
 \\
 & CC
     & \cellcolor{red!100}\bf0.00
     & \cellcolor{red!76}\bf0.24
     & \cellcolor{red!100}\bf0.00
     & \cellcolor{green!15}\bf1.51
     & \cellcolor{red!55}\bf0.44
     & \cellcolor{red!86}\bf0.14
     & \cellcolor{red!90}\bf0.09
 \\
 & BFS
     & \cellcolor{green!46}\bf4.61
     & \cellcolor{green!14}\bf1.48
     & \cellcolor{red!100}\bf0.00
     & \cellcolor{green!35}\bf3.59
     & \cellcolor{red!20}\bf0.80
     & \cellcolor{green!34}\bf3.45
     & \cellcolor{green!57}\bf5.73
 \\
 & TC
     & \cellcolor{red!100}\bf0.00
     & \cellcolor{red!26}\bf0.74
     & \cellcolor{red!100}\bf0.00
     & \cellcolor{red!95}\bf0.04
     & \cellcolor{red!98}\bf0.02
     & \cellcolor{red!70}\bf0.29
     & \cellcolor{red!77}\bf0.23
 \\
 \hline
 \multicolumn{9}{c}{} \\
 \hline
 \parbox[t]{2mm}{\multirow{5}{*}{\rotatebox[origin=c]{90}{\bf PGAbB-GPU}}}
  & PR
      & \cellcolor{green!45}\bf4.20
      & \cellcolor{green!46}\bf4.72
      & \cellcolor{red!36}\bf0.74
      & \cellcolor{red!47}\bf0.53
      & \cellcolor{red!35}\bf0.64
      & \cellcolor{green!100}\bf13.60
      & \cellcolor{green!23}\bf2.30
  \\
  & SV/LP
      & \cellcolor{green!100}\bf19.19
      & \cellcolor{green!99}\bf9.96
      & \cellcolor{green!31}\bf3.16
      & \cellcolor{green!64}\bf6.45
      & \cellcolor{green!36}\bf3.63
      & \cellcolor{green!92}\bf9.21
      & \cellcolor{green!38}\bf3.85
  \\
  & CC
      & \cellcolor{green!16}\bf1.68
      & \cellcolor{green!10}\bf1.08
      & \cellcolor{green!55}\bf5.52
      & \cellcolor{green!35}\bf3.56
      & \cellcolor{green!13}\bf1.37
      & \cellcolor{red!35}\bf0.64
      & \cellcolor{red!69}\bf0.31
  \\
  & BFS
      & \cellcolor{red!82}\bf0.18
      & \cellcolor{red!15}\bf0.85
      & \cellcolor{red!3}\bf0.97
      & \cellcolor{red!72}\bf0.28
      & \cellcolor{red!68}\bf0.32
      & \cellcolor{green!10}\bf1.06
      & \cellcolor{red!73}\bf0.27
  \\
  & TC
      & \cellcolor{green!30}\bf3.09
      & \cellcolor{green!33}\bf3.39
      & \cellcolor{green!23}\bf2.34
      & \cellcolor{red!47}\bf0.52
      & \cellcolor{red!67}\bf0.32
      & \cellcolor{green!28}\bf2.87
      & \cellcolor{green!23}\bf2.33
  \\ 
\hline
\hline
\parbox[t]{2mm}{\multirow{5}{*}{\rotatebox[origin=c]{90}{\bf PGAbB}}}
 & PR
     & \cellcolor{green!46}\bf4.64
     & \cellcolor{green!46}\bf4.67
     & \cellcolor{red!19}\bf0.80
     & \cellcolor{red!47}\bf0.53
     & \cellcolor{red!36}\bf0.64
     & \cellcolor{green!100}\bf10.76
     & \cellcolor{green!17}\bf1.79
 \\
 & SV/LP
     & \cellcolor{green!100}\bf18.02
     & \cellcolor{green!59}\bf5.95
     & \cellcolor{green!19}\bf1.90
     & \cellcolor{green!57}\bf5.73
     & \cellcolor{green!29}\bf2.95
     & \cellcolor{green!76}\bf7.70
     & \cellcolor{green!19}\bf1.98
 \\
 & CC
     & \cellcolor{green!12}\bf1.25
     & \cellcolor{green!15}\bf1.53
     & \cellcolor{green!21}\bf2.14
     & \cellcolor{green!19}\bf1.91
     & \cellcolor{red!3}\bf0.96
     & \cellcolor{green!23}\bf2.40
     & \cellcolor{red!13}\bf0.87
 \\
 & BFS
     & \cellcolor{red!83}\bf0.16
     & \cellcolor{red!11}\bf0.89
     & \cellcolor{red!23}\bf0.77
     & \cellcolor{red!9}\bf0.90
     & \cellcolor{red!67}\bf0.33
     & 1.00
     & \cellcolor{red!71}\bf0.29
 \\
 & TC
     & \cellcolor{green!30}\bf3.02
     & \cellcolor{green!30}\bf3.01
     & \cellcolor{green!16}\bf1.69
     & \cellcolor{green!11}\bf1.11
     & \cellcolor{green!39}\bf3.91
     & \cellcolor{green!53}\bf5.39
     & \cellcolor{green!34}\bf3.48
 \\
\hline
\end{tabular}
\end{center}
\label{tab:multicol}
\end{table*}

\subsection{Dataset}
\label{ssec:dataset}

To include different kind of medium size graphs;
we downloaded a total of 44 graphs that have edges from 100 millions to 2.1
billion from SuiteSparse\footnote{SuiteSparse: {https://suitesparse-collection-website.herokuapp.com}},
Konect\footnote{Konect: {http://konect.cc/networks/}}, and
Snap\footnote{Snap: {http://snap.stanford.edu/data/index.html}} dataset
repositories. We transformed all graphs to undirected, and removed duplicate
edges.

Due to space limitations, we selected seven
different types of graphs from our dataset for detailed results; two social networks,
two synthetic graphs, a road network, a gene graph, and a web graph.
In \pgabb selected seven graphs can fit into GPU memory. Therefore
we also included \pgabb's GPU-only performance on those graphs.
Note that, to process larger problem sizes we need heterogeneous execution
model of \pgabb.
We are
going to use GAPBS as the reference implementation in our comparisons.
Table~\ref{tab:multicol} groups performances of frameworks on five algorithms.
Each cell presents a framework's speedup on a graph and a graph algorithm.
In a particular case the value of a cell is:
(1) $1.0$ if the framework performs the same as GAPBS.
(2) $2.0$ if the framework performs two times faster than GAPBS.
(3) $0.5$ if the framework performs two times slower than GAPBS.
We use a heat map to visually represent performances. In recent work,
Azad et al.~\cite{Azad20-IISWC} provide a similar table to present the results
of different frameworks. Our findings match that work.

\subsection{Evaluation of bulk graph computations}

\subsubsection{PageRank algorithm}

There exist several graph algorithms that follow SpMV (Sparse matrix-vector
multiplication) type of computation. PageRank is one of the most popular ones.
In this experiment, we used SpMV like PageRank implementations.
We choose damping factor as $0.85$, error tolerance as $0.0001$, and iteration
limit as $20$. We report the average execution time of a PageRank iteration.
Gunrock and Galois-GPU fail to process twitter7 and sk-2005 graphs due to GPU
memory limitations.

\pgabb gives the best performance in twitter7, Orkut, and kron21 graphs, and
the second performance on the myciel19 graph.
Atomic updates becomes the bottleneck on sk-2005, kmer\_V1r and eu\_osm
graphs. Because, in our PageRank algorithm implementation we use atomic
operations to update ranks of the vertices even in the pull direction due to
two-dimensional block layout. sk-2005 graphs high locality, kmer\_V1r and
eu\_osm graphs very high low degree vertices ($\approx99\%$ of the vertices
have less than two neighbors) increases the cost of atomic operations
drastically.
We observe that when graphs fit into GPU memory \pgabb's GPU-only
execution performs slightly better than hybrid execution. Bandwidth
between CPU and GPU is the primary factor of this. The gain comes from
CPUs cannot compensate the synchronization cost of the arrays between
GPU and CPU.

PageRank algorithm implementations of these systems are very competitive.
On the complete dataset; in median, \pgabb performs
$2.4\times$, $1.3\times$, $1.8\times$, $1.7\times$, $1.1\times$, and
$1.6\times$ better than GAPBS, Galois, Ligra, LAGraph Galois-GPU, and Gunrock,
respectively.

\subsubsection{Shiloach-Vishkin or Label Propagation algorithms}

In this experiment, we are going to evaluate performances of SV
(Shiloach-Vishkin) or LP (Label Propagation) algorithm implementations in
different systems. Because, some systems provide an implementation of SV
algorithm and some LP algorithm. Both algorithms have similar execution
principles. The main difference between those two algorithms is the
upper bound of the number of iterations: SV algorithm's iteration bound
is $O(\log(|V|))$ and LP algorithms iteration bound is $O(D)$ where $D$
is the diameter of the graph. SV algorithm's compression step causes this
difference. Gunrock and Ligra implement LP algorithm. The other
systems implement SV algorithm. LAGraph implements an optimized SV
algorithm~\cite{zhang20-SIAMPP} which can converge faster on some graphs.

To achieve better performance, \pgabb executes hooking step in the GPU and
linking step in CPUs. Between those steps \pgabb synchronizes the component
array (a global attribute). That synchronization cost is included in the
reported execution times.

Galois-GPU and Gunrock fail to process the largest twitter7 and sk-2005
graphs, due to device memory limitations.  \pgabb performs the best.
Galois, and Gunrock have the second and the third best performances.
Ligra's poor performance has two reasons; more number of iterations (LP
algorithm) and imbalance between computational loads. OpenMP might be the
cause of the latter reason, because Ligra is primarily designed using Cilk.
Similar to PageRank algorithm, we observe that when graphs fit into
GPU memory \pgabb's GPU-only execution performs better than hybrid
execution. Synchronization between CPU and GPU becomes the bottleneck.

On the complete dataset; in median, \pgabb performs $3.9\times$, $2.0\times$,
$20.0\times$, $3.9\times$, $1.4\times$, and $1.9\times$ better than
GAPBS, Galois, Ligra, LAGraph Galois-GPU, and Gunrock respectively.

\subsubsection{Best connected-components algorithm}

In this experiment, we are going to evaluate performances of best performing
connected component algorithm implementations in different systems.
GAPBS, Galois and \pgabb implement Afforest
algorithm~\cite{Sutton18-IPDPS}. Ligra implements low-diameter graph
decomposition based algorithm~\cite{Shun14-SPAA}. The other systems
do not have a specific implementation and implement SV or LP algorithms.

To achieve better performance, \pgabb executes sampling step in the GPU and
finalization step in CPUs. Again, synchronization cost is included in the
reported execution times.

Galois-GPU and Gunrock fail to process the largest twitter7 and sk-2005
graphs. We observe that
Afforest algorithm's lower computational complexity is the main advantage of
GAPBS, Galois and \pgabb. Even though Ligra's implementation is
work-efficient it requires more read/write operations and perform poor. Ligra
suffers from bad workload imbalance again.
Besides of eu\_osm and kron21 graphs, \pgabb's implementation outperforms
Galois, and GAPBS.
Similar to previous algorithms, on majority of the
graphs \pgabb's GPU-only execution performs better than hybrid
execution due to synchronization between CPU and GPU. However, we also
observe that on more irregular graphs with many components (mycielskian19 and
kron21) GPU-only execution performs worse than hybrid execution. Because,
CPUs handle linking operation more efficiently and the gain
compansates the synchronization cost.

On the complete dataset; in median, \pgabb performs $1.5\times$, $1.4\times$,
$105\times$, $11.2\times$, $3.7\times$, and $4.7\times$ better than
GAPBS, Galois, Ligra, LAGraph Galois-GPU, and Gunrock respectively.

\subsection{Evaluation of traversal computations}

In this experiment, we are going to evaluate performances of different systems
on a traversal graph algorithm; direction-optimized BFS
algorithm~\cite{Beamer12-SC}. Galois-GPU hanged and could not process those
graphs in our machine. Gunrock failed to process the sk-2005 graph due to
device memory limitations.

Gunrock performs the best. The others have similar performances, but with
slightly better execution times Ligra and \pgabb gives the second and third
best performances. Gunrock's design is a perfect fit for the BFS problem and
it leverages massive parallelization capabilities of GPUs. As the runtime is
small, the overheads of graph processing systems are significant in the final
performance even if they implement the same algorithm as GAPBS therefore we
observe a significant slow down on the other systems. In addition,
depending on the graph structure, two-dimensional layout also brings
additional cost to \pgabb. In an adversarial case \pgabb might require to
visit all edges and cannot benefit from the direction optimization.
We observe that \pgabb's GPU-only and hybrid execution are close.

On the complete dataset; in median, \pgabb performs
$1.6\times$, $1.2\times$, and $3.1\times$, worse than GAPBS, Ligra
and Gunrock, respectively. \pgabb performs $1.7\times$, and $1.1\times$
better than Galois and LAGraph, respectively.

\subsection{Evaluation of pattern-based computations}

In this experiment, we are going to evaluate performances of different systems
on a non-trivial graph algorithm; triangle counting.
Degree-based vertex ordering is a commonly used
heuristic in triangle counting algorithms. Therefore we enabled degree
ordering in all systems. We do not report time spent for this process.

Gunrock fails to process twitter7 and sk-2005 graphs.
Triangle counting algorithm only requires half of the edges, therefore
Galois-GPU was able to process all graphs.
Thanks to design flexibilities \pgabb performs the best. GAPBS, and Galois
have the second and the third best performances. Workload imbalance becomes
the bottleneck for Ligra.
\pgabb's GPU-only execution is more fragile to the graph structure. Because,
sparse tasks are more bandwidth bounded than the denser tasks and takes more
time on GPUs.
We observe that on this problem hybrid execution handles the performance
variety caused by the very sparse tasks more successful by assigning
them to CPUs.

On the complete dataset; in median, \pgabb performs $1.7\times$, $2.6\times$,
$4.9\times$, $2.9\times$, $5.9\times$, and $10.5\times$ better than
GAPBS, Galois, Ligra, LAGraph Galois-GPU, and Gunrock respectively.

\subsection{Overall results}

\begin{figure}[htb]
  \centering
  \includegraphics[width=.75\linewidth]{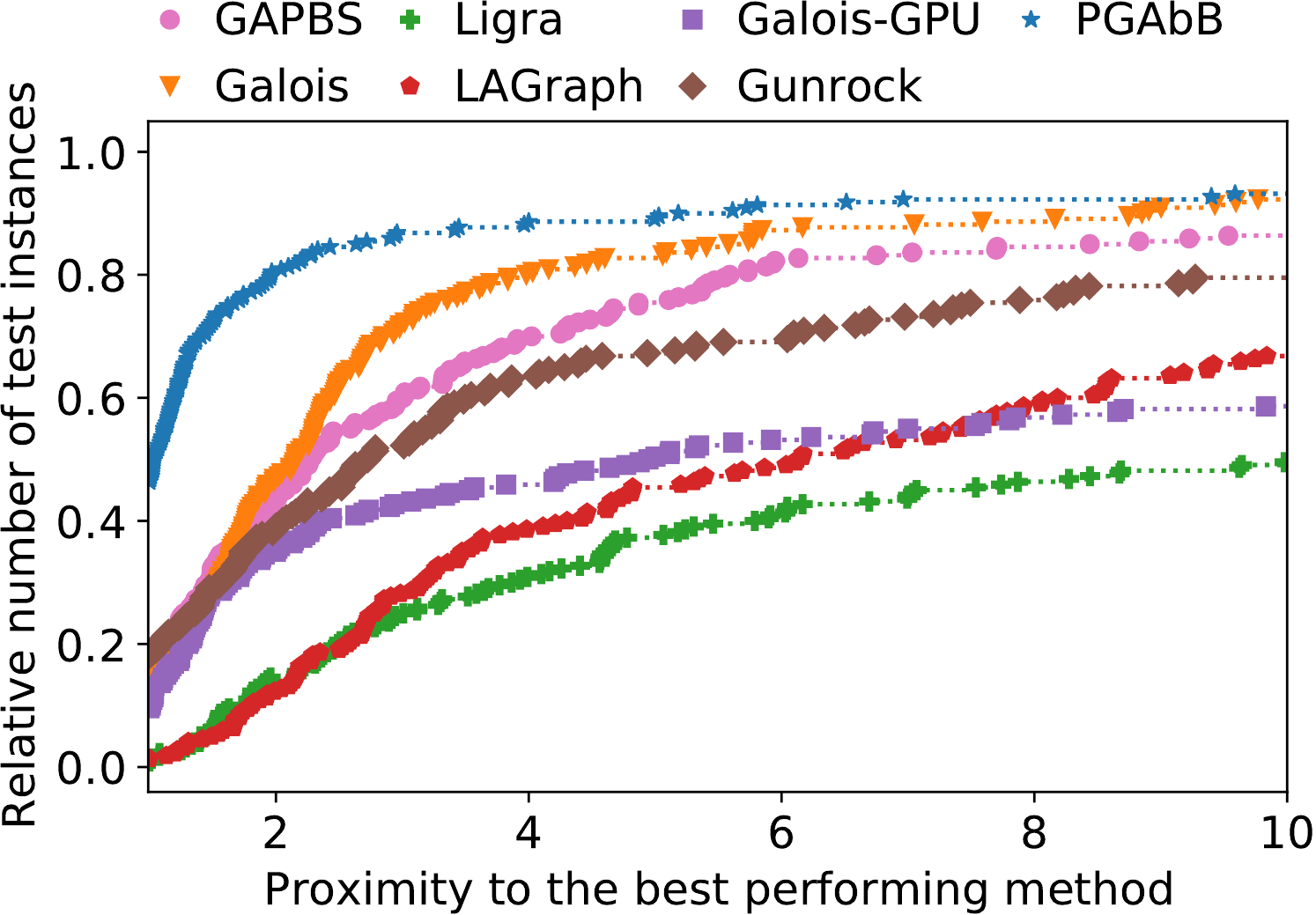}
  \caption{Performance profiles of all combined implementations. Higher and
  closer to the $y$-axis: system is better.}
  \label{fig:all}
\end{figure}

Fig.~\ref{fig:all} illustrates performance profiles~\cite{Dolan02-MP}
of seven systems on
complete $220$ tests; $5$ algorithms on $44$ graphs. In $\approx 46\%$ of the
test instances \pgabb gives the best performance. Considering overall
performances of those seven systems we can subjectively rank them
as follows: $1^{\text{st}}$ \pgabb, $2^{\text{nd}}$ Galois, $3^{\text{rd}}$
GAPBS, $4^{\text{th}}$ Gunrock, $5^{\text{th}}$ Galois-GPU, $6^{\text{th}}$ LAGraph,
and $7^{\text{th}}$ Ligra.
In the median, \pgabb performs $1.6\times$, $1.6\times$,
$5.7\times$, $3.4\times$, $4.5\times$, and $2.4\times$ better than
GAPBS, Galois, Ligra, LAGraph Galois-GPU, and Gunrock respectively.

\section{Conclusion}
\label{sec:conc}

This paper introduces \pgabb: a block-based algorithmic framework for parallel
graph processing on heterogeneous platforms. In a heterogeneous setting,
\pgabb aims to maximally leverage from different architectures by implementing
a task-based execution on top of a block-based programming model. Our
experimental results show that  in the median, \pgabb performs $1.6\times$ to
$5.7\times$ better than state-of-the-art four CPU-based and two GPU-based graph
processing frameworks over $44$ graphs in the range of $100$M to $2.1$B edges.

As a future work, our goal is to make \pgabb
distributed by integrating locality-aware schedulers and garbage collectors.
In addition, we plan to investigate, developing automatic parameter tuning
approaches that will take into account properties of the graph, 
the kernel, and compute architectures.

\vspace{1em}
\noindent {\bf Acknowledgments:}
This work was supported in parts by the NSF grants CCF-1919021 and Sandia
National Laboratories.
Sandia National Laboratories is a multi-mission laboratory managed and
operated by National Technology and Engineering Solutions of Sandia,
LLC., a wholly owned subsidiary of Honeywell International, Inc., for
the U.S. Department of Energy's National Nuclear Security Administration
under contract DE-NA-0003525.

\bibliographystyle{IEEEtran}
\bibliography{paper,tdalab}

\begin{IEEEbiography}[{\includegraphics[width=1in,clip,keepaspectratio]{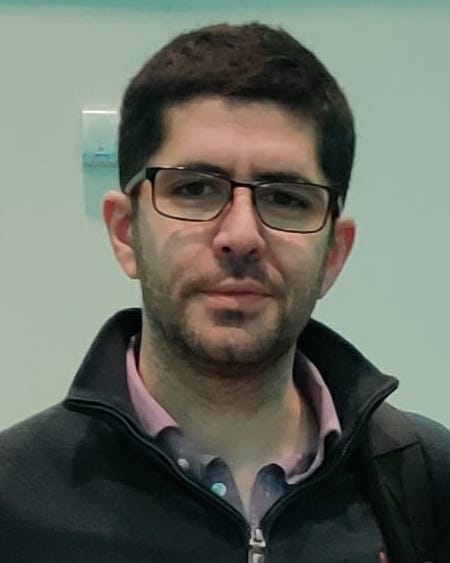}}]{Abdurrahman Ya\c{s}ar}
is a Compute Developer Technology Engineer at Nvidia. He holds a Ph.D. in 
Computer Science from the School of Computational Science and Engineering at
Georgia Institute of Technology. He received his M.S. in Computer Engineering
from Bilkent University, Turkey in 2015.
\end{IEEEbiography}

\begin{IEEEbiography}[{\includegraphics[width=1in,clip,keepaspectratio]{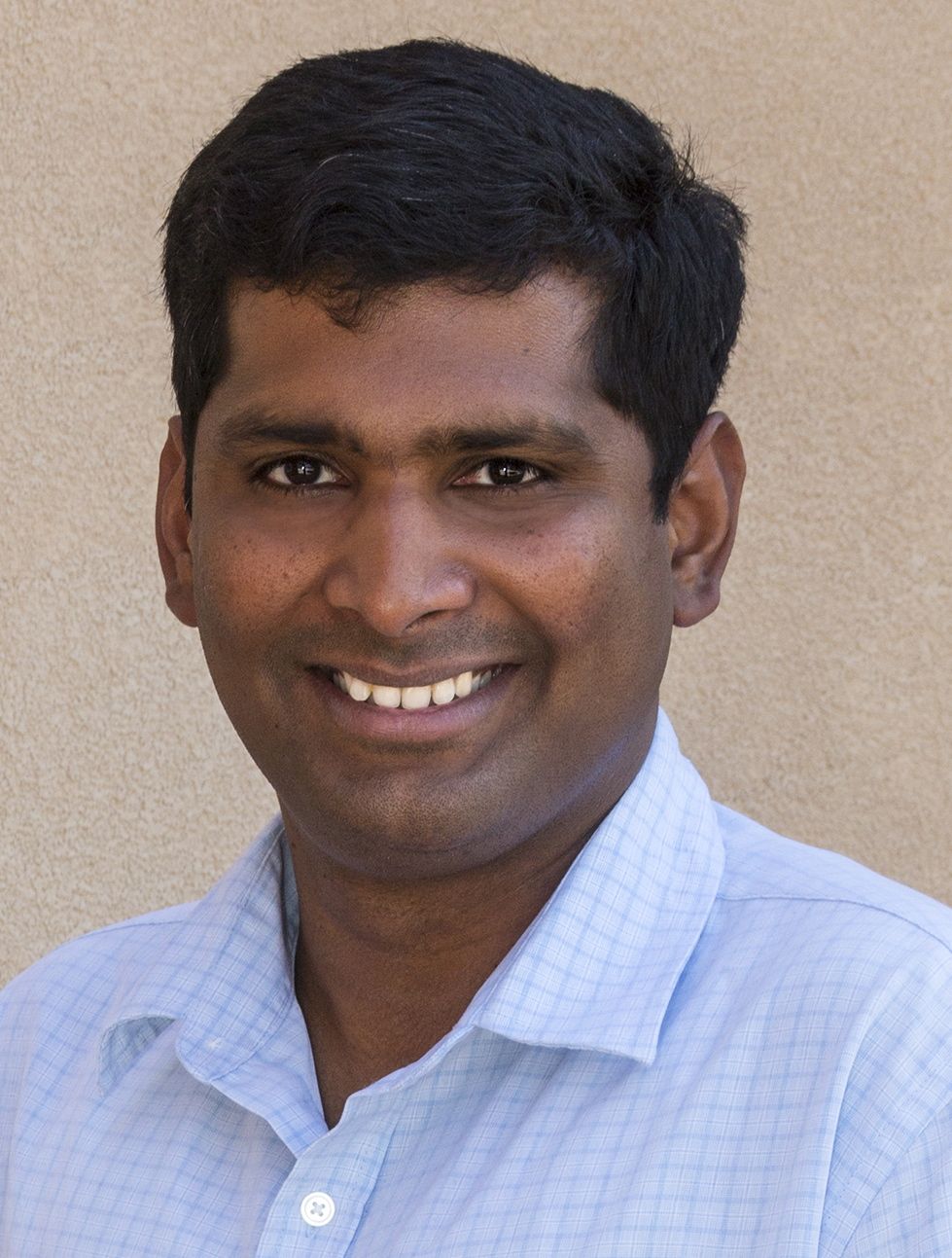}}]{Sivasankaran Rajamanickam}
(M’14) is a Principal Member of Technical Staff in the Center for Computing
Research at Sandia National Laboratories. He earned his B.E. from Madurai
Kamaraj University, India, and his Ph.D. in Computer Engineering from University
of Florida.
\end{IEEEbiography}

\begin{IEEEbiography}[{\includegraphics[width=1in,clip,keepaspectratio]{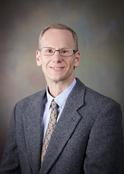}}]{Jonathan W. Berry}
is a Distinguished Member of the Technical Staff at Sandia National
Laboratories.  He holds a Ph.D. in computer science from Rensselaer Polytechnic
Institute and spent almost a decade in liberal arts academia before joining
Sandia in 2004.
\end{IEEEbiography}

\begin{IEEEbiography}[{\includegraphics[width=1in,clip,keepaspectratio]{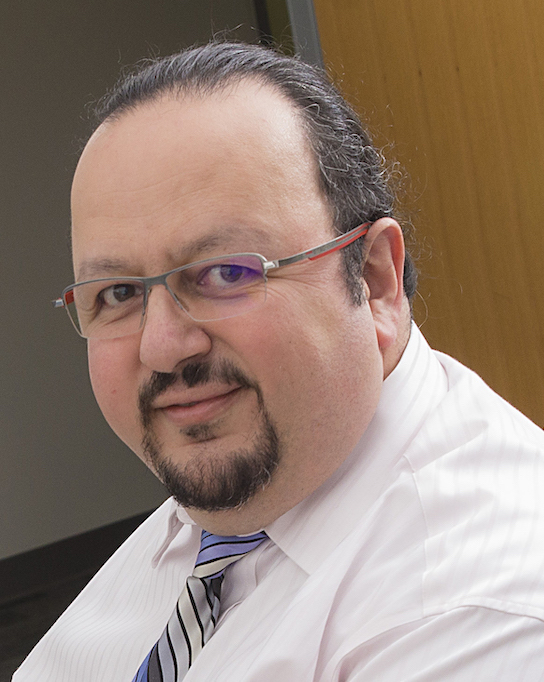}}]{\"Umit V. \c{C}ataly\"urek}
(M’09,SM’10,Fellow’16) is a Professor and Associate Chair in the School of
Computational Science and Engineering at the Georgia Institute of Technology. He
received his Ph.D., M.S. and B.S. in Computer Engineering and Information
Science from Bilkent University, Turkey.
\end{IEEEbiography}

\end{document}